\documentclass[10pt,a4paper,onecolumn,twoside]{article}
\usepackage[T1]{fontenc}
\usepackage[utf8]{inputenc} 
\usepackage[english]{babel}
\usepackage{textcomp}
\usepackage{amsmath}
\usepackage{amsthm} 
\usepackage{amssymb}
\usepackage[top=3.0cm, bottom=2.0cm, inner=2.0cm, outer=3.0cm]{geometry}
\usepackage{graphicx}
\usepackage{color}
\usepackage[hidelinks]{hyperref}

\newcounter{SaveEqnCntr}
\newcounter{SaveEqnCntrRom}

\newtheorem{defn}{\textbf{Definition}}
\newtheorem{thm}{\textbf{Theorem}}

\newtheorem{prop}{\textbf{Proposition}}
\newtheorem{rem}{\textbf{Remark}}

\title{A small-gain theorem for 2-contraction of nonlinear interconnected systems}
\author{David Angeli, Davide Martini, Giacomo Innocenti and Alberto Tesi
\thanks{David Angeli is with the Department of Electrical and Electronic Engineering, Imperial College London, London, SW7 2AZ, UK. {\textit{ d.angeli@imperial.ac.uk}} and with the Dept of Information Engineering of University of Florence; Giacomo Innocenti, Davide Martini and Alberto Tesi are with the Dept of Information Engineering of University of Florence.}
}
\date{}

\begin{document}

\maketitle

\begin{abstract}
This paper introduces small-gain sufficient conditions for $2$-contraction of feedback interconnected systems, on the basis of individual gains
of suitable subsystems arising from a modular decomposition of the second additive compound equation. The condition applies even to cases when individual subsystems might fail to be contractive (due to the extra margin of contraction afforded by the second additive compound matrix).
Examples of application are provided to illustrate the theory and show its degree of conservatism and scope of applicability. 
\end{abstract}

\section{Introduction}
The study of stability and convergence properties for nonlinear dynamical systems is a classical topic that is usually traced back to the work of Lyapunov, (see \cite{lyapunov} for a translation in English). From its origin, two complementary approaches have been pursued, viz. the so called direct method, involving candidate Lyapunov functions, or the indirect method, based for a differentiable vector field $f(x)$ on the consideration of linearized dynamics, captured by the Jacobian $\frac{\partial f}{\partial x}$.\\
While the indirect method was initially confined as a local analysis tool, specifically for equilibrium solutions, several generalisations have emerged in the subsequent decades, 
in particular extending the approach to periodic solutions~\cite{floquet}, to complex regimes~\cite{oseledets}, and also to regional or global results (see~\cite{jouffroy,Pavlov} and references therein for an historical account of such extensions).

A renewed interest in the subject was triggered by the seminal paper~\cite{slotinecontraction}, which established the name of Contraction Theory for the area of the stability analysis based on the use of variational equations, also interpreted in terms of virtual displacements and their extensions.   
Rather than monitoring the evolution of state perturbations along any particular nominal solution of interest, contraction analysis directly postulates sufficient stability criteria within prespecified (typically forward invariant) ``contraction regions'' by assuming a certain Linear Matrix Inequality (LMI) condition, which entails convergence of solutions towards each other, i.e.:
\[  \frac{\partial f}{\partial x}^T (x) P(x) + P(x) \frac{ \partial f}{\partial x} (x) + \dot{P}(x)<0,\]
for some symmetric, positive definite and differentiable matrix function $P(x)$ of the system state $x$.
Such conditions typically entail a stronger form of stability, also referred to in the literature as Incremental Stability~\cite{angeliincremental}. Indeed, connections between the Lyapunov direct method and contraction analysis have gradually consolidated and they were recently extended in~\cite{fornidifferential}.

Several interesting extensions of contraction analysis have emerged in recent years. For instance, \cite{manchester} deals with contraction analysis of periodic solutions, while~\cite{diffpositive} introduces LMIs where the typical positive definiteness requirement on the matrix $P(x)$ is relaxed to an inertia constraint on the spectrum of $P(x)$ (a single negative eigenvalue), while at the same time the focus of the analysis is shifted to ruling out limit cycles and to still enforcing convergence to equilibrium solutions, which is of great interest in many applications, where, for instance, multistable dynamical behaviors are allowed and aimed for.
A related approach, again devoted at ruling out existence of periodic solutions, was proposed several decades ago in a remarkable paper by James Muldowney~\cite{muldowney}. This paper introduces the use of compound matrices in the study of linear and nonlinear differential equations. In particular, what would be interpreted in today's language as a contraction assumption on the second additive compound matrix of the Jacobian, was shown to forbid existence of periodic solutions.

Such results have caught the attention of the scientific community and, in recent years, have motivated the introduction of the so called $k$-Contraction Theory, in rough terms contraction of arbitrary virtual parallelotopic displacements of dimension $k$~\cite{kcontraction}.
See also~\cite{margaliotcompound} for a recent survey on the topic. In such theory, the case $k=2$ plays a special role, as it corresponds to the conditions introduced in~\cite{muldowney}, which have a strong direct link with the dynamics of the original nonlinear system.

In this respect, \cite{muradangeli} used non-singularity of the second additive compound matrix (verified through suitable associated graphs) to structurally rule out existence of Hopf's bifurcations in Chemical Reaction Networks. Similarly, motivated by the study of biological interaction networks, \cite{muhammadIEEE} formulated Lyapunov-based conditions and contraction criteria on the second additive compound matrix of the Jacobian for ruling out periodic and almost periodic solutions.  Slightly relaxed conditions are used in~\cite{IEEEmartini} to rule out positive Lyapunov exponents in non-equilibrium attractors found within prescribed forward invariant regions of the state space. A related line of investigation generalises compound matrices to non-integer orders so to provide constructive criteria for the estimation of the Hausdorff dimension of attractors in nonlinear dynamical systems~\cite{margaliotslotine}. 

At the same time, stability analysis of interconnected dynamical systems has also become a very active area of research. The general idea behind a Small-Gain Theorem is the formulation of a sufficient stability condition, for a feedback interconnected system of some sort, on the basis of the stability of its modular components, and the calculation of some notion of ``loop gain,'' which, if sufficiently low (typically smaller than unity), is adequate for assessing the stability of the whole interconnection. Many versions of such result exist, ranging from Input-to-State-Stability (ISS) systems~\cite{sgt} to an LMI set-up~\cite{lmisgt}, and passing for large-scale interconnected systems~\cite{largescale}. See~\cite{bullobook} for a recent and up-to-date reference, where modular techniques for contraction analysis of large-scale networks are perfected and treated in depth.  

The special case of $k$-contraction for two cascaded systems is studied in~\cite{kseries}, while in~\cite{margaliotlurie} the case of static nonlinear feedback (of the Lurie form) is considered.\\ 
In the present note, we formulate a small-gain theorem result for $2$-contraction of feedback interconnected systems, based on the second additive compound matrix of individual subsystems and of an auxiliary coupling systems, which captures the dynamics of their interconnections.
In particular, rather than resolving a unique LMI condition of size ${n \choose 2} \times {n \choose 2}$, we consider subsystems of dimension $n_1$ and $n_2$ (with $n_1+n_2=n$) and we solve $3$ separate LMIs with unknowns of size ${n_1 \choose 2}$, ${n_2 \choose 2}$ and $n_1 \cdot n_2$ respectively.\\
The rest of the paper is organised as follows: Section 2 introduces the key definitions and preliminary results for a modular formulation of the $2$-contraction property; Section 3 formulates suitable notions of gains for linear systems through the use of LMIs and it proposes a first small-gain theorem; Section 4 extends the technique to the case of nonlinear systems and state-dependent contraction metrics; Section 5 proposes examples of applications to illustrate the theory and its conservatism; Section 6 draws some conclusions and future research directions.

\section{Problem formulation and preliminary results}
\label{sec:prelim}
Consider for the time being an interconnected linear system of the following form:
\begin{equation}
\label{interconnected}
 \left [ \begin{array}{c}\dot{x}_1 \\ \dot{x}_2 \end{array} \right ]
 = \left [ \begin{array}{cc} A_{11} & A_{12} \\ A_{21} & A_{22}   \end{array} \right ] \, \left [ \begin{array}{c} x_1 \\ x_2 \end{array} \right ], 
\end{equation}
where $x_1$ and $x_2$ are vectors of dimension $n_1, n_2 \geq 2$, and $A_{11}$, $A_{12}$, $A_{21}$ and $A_{22}$
are blocks of compatible dimensions, with $A_{11}$ and $A_{22}$ being square.
We interpret equation (\ref{interconnected}) as the equation of a feedback interconnection of the $x_1$ and $x_2$ sub-systems. In particular, off-diagonal blocks may be of low rank, (corresponding to fewer input and output variables), but this is not needed for the results to follow.
We denote by $A$ the block-matrix
\begin{align}
	A = \left [ \begin{array}{cc} A_{11} & A_{12} \\ A_{21} & A_{22}   \end{array} \right ],
\label{eq:partitioned A}
\end{align}
and are interested in modular conditions to guarantee asymptotic stability of the second additive compound matrix
$A^{[2]}$.
In the following, two distinct operators are needed to convert matrices into vectors.
In particular, for a $n \times n$ skew-symmetric matrix $X$ we denote by:
\begin{align*} 
	\vec{X} = [x_{12}, x_{13}, \ldots, x_{1n}, x_{23}, x_{24}, \ldots, x_{2n}, \ldots, x_{(n-1)n} ]^T. 
\end{align*}
Instead, for a $m \times n$ rectangular matrix $X$ we denote by:
\begin{align}
\textrm{vec}(X) = [x_{11},x_{12}, \ldots, x_{1n}, x_{21},x_{22}, \ldots, x_{2n}, \ldots, x_{m1}, \ldots, x_{mn}]^T.
\label{vec2}
\end{align}
Notice that there exists a matrix $M_n \in \mathbb{R}^{n^2 \times { n \choose 2 }}$ such that for any skew-symmetric matrix $X \in \mathbb{R}^{n \times n}$, it holds
\begin{equation}
\label{conversion}
 \textrm{vec}(X) = M_n \vec{X}.   
\end{equation}
In particular, $M_n$ is given as:
\[ M_n = \sum_{1 \leq i \neq j \leq n} \textrm{sign}(j-i) e_{[(i-1)n + j]} e_{k(i,j)}^T  \]
where 
\[ k(i,j) = |i-j|+ { n \choose 2 } - { {n+1 - \min \{i,j \}} \choose 2}. \]
For clarity, matrix $M_4$ is shown next:
\[ M_4 = \left [ \begin{array}{cccccc} 0 & 0 & 0 & 0 & 0 & 0 \\
1 & 0 & 0& 0 & 0 & 0 \\
0 & 1 & 0 & 0 & 0 & 0 \\
0 & 0 & 1 & 0 & 0 & 0 \\
-1& 0 & 0 & 0 & 0 & 0 \\
0 & 0 & 0 & 0 & 0 & 0 \\
0 & 0 & 0 & 1 & 0 & 0 \\
0 & 0 & 0 & 0 & 1 & 0 \\
0 & -1 & 0 & 0 & 0 & 0 \\
0 & 0 & -1 & 0 & 0 & 0 \\
0 & 0 & 0 & 0 & 0 & 0 \\
0 & 0 & 0 & 0 & 0 & 1 \\
0 & 0 & 0 & -1 & 0 & 0 \\
0 & 0 & 0 & 0 & -1 & 0 \\
0 & 0 & 0 & 0 & 0 & -1 \\
0 & 0 & 0 & 0 & 0 & 0
\end{array} \right ] \]
Conversely, there exists a matrix $L_n \in \mathbb{R}^{{ n \choose 2 } \times n^2}$ such that  for any skew-symmetric matrix $X \in \mathbb{R}^{n \times n}$ the following holds:
\begin{equation}
 \label{conversion2}
 \vec{X} = L_n \textrm{vec} ( X ),
\end{equation}
where $L_n$ is given as:
\[ L_n = \sum_{1 \leq i < j \leq n}  e_{k(i,j)} e_{[(i-1)n + j]}^T.  \]
As an example, $L_4$ is reported hereafter:
\begin{align}
		L_4 = \left  [ \begin{array}{cccccccccccccccc} 0 & 1 & 0 & 0 & 0 & 0 & 0 & 0 & 0 & 0 & 0 & 0 & 0 & 0 & 0 & 0 \\
                                                 0 & 0 & 1 & 0 & 0 & 0 & 0 & 0 & 0 & 0 & 0 & 0 & 0 & 0 & 0 & 0 \\
                                                 0 & 0 & 0 & 1 & 0 & 0 & 0 & 0 & 0 & 0 & 0 & 0 & 0 & 0 & 0 & 0 \\
                                                 0 & 0 & 0 & 0 & 0 & 0 & 1 & 0 & 0 & 0 & 0 & 0 & 0 & 0 & 0 & 0 \\
                                                 0 & 0 & 0 & 0 & 0 & 0 & 0 & 1 & 0 & 0 & 0 & 0 & 0 & 0 & 0 & 0 \\
                                                 0 & 0 & 0 & 0 & 0 & 0 & 0 & 0 & 0 & 0 & 0 & 1 & 0 & 0 & 0 & 0
                                                 \end{array} \right ]
\label{matriceL4}
\end{align}

Skew-symmetric matrices are useful in this context due to the following result linking them to second additive compound matrices.
Assume that a given skew-symmetric matrix function $X$ fulfills the matrix differential equation:
\begin{equation}
\label{skewequation}
  \dot{X} = AX + X A^T.  
\end{equation}   
It is easy to verify that the linear operator $L(X) = AX + XA^T$ preserves skew symmetry.
In particular, $L(X)^T=-L(X)$ for all skew-symmetric $X$.
Moreover, it is known that the vector $\vec{X}$ fulfills the differential equation:
\begin{equation}
    \label{coupled}
	\dot{\vec{X} } = A^{[2]} \vec{X}, 
\end{equation}
where $A^{[2]}$ is the second additive compound matrix of $A$~\cite{linearmarcus,muradangeli}. 
Consider next a skew-symmetric matrix $X$, which is partitioned according to $A$ as in~(\ref{eq:partitioned A}), i.e.,
\begin{equation}
\label{xpartition}
     X= \left [ \begin{array}{cc} X_{11} & X_{12} \\ X_{21} & X_{22} \end{array} \right ].
\end{equation} 
By skew-symmetry we have that $X_{11}^T=-X_{11}$ and $X_{22}^T=-X_{22}$, viz. diagonal blocks are themselves skew-symmetric. In addition, $X_{21}T=-X_{12}$.
Our goal is to decompose the dynamics of (\ref{coupled}) by looking at the different state-components $\vec{X}_{11}$, $\vec{X}_{22}$ and $\textrm{vec} ( X_{12} )$.
Our first result is the following.
\begin{prop} Consider the matrix-valued differential equation (\ref{skewequation}), and assume that its unknown $X$ is a skew-symmetric matrix partitioned according to (\ref{xpartition}).
Then, the vectors $\vec{X}_{11}$, $\vec{X}_{22}$ and $\textrm{vec} ( X_{12} )$ fulfill the following linear system of coupled differential equations:
\begin{align}
	\left\{\begin{array}{l}
	\dot{\vec{X}}_{11} = A_{11}^{[2]} \vec{X}_{11} + B_1 \textrm{vec}(X_{12})  \\
	\dot{\vec{X}}_{22} = A_{22}^{[2]} \vec{X}_{22} + B_2 \textrm{vec}(X_{12})  \\
 	\textrm{vec} (\dot{X}_{12})  = ( A_{11} \oplus A_{22} )  \textrm{vec} (X_{12})
+ G_1 \vec{X}_{11} + G_2 \vec{X}_{22}
	\end{array}\right.
	\label{coupledsystem}
\end{align}
where the matrices $B_1$, $B_2$, $G_1$ and $G_2$ are given by:
\begin{align*}
	B_1 &=  L_{n_1} ( I_{n_1} \otimes A_{12} ) - L_{n_1} ( A_{12} \otimes I_{n_1} ) H_{n_1,n_2} \\
	B_2 &= L_{n_2} ( I_{n_2} \otimes A_{21} ) - L_{n_2} ( A_{21} \otimes I_{n_2} ) H_{n_2,n_1} \\
	G_1 &= (I \otimes A_{21} ) M_{n_1} \\
	G_2 &= (A_{12} \otimes I) M_{n_2}
\end{align*}
and the matrix $H_{n_1,n_2}$ is defined as
\[ H_{n_1,n_2} = \sum_{i=1}^{n_1} \sum_{j=1}^{n_2} e_{[(j-1)n_1 + i]} e_{[(i-1) n_2 + j]}^T  ~,\] 
converts row vectorisation to column vectorisation, viz. $\textrm{vec} ( X_{12}^T ) = H_{n_1,n_2} \textrm{vec} (X_{12} )$.
\end{prop}
\emph{Proof.} To see the result, compute the block-partitioned expression of $\dot{X}$ according to:
\begin{align*} 
\dot{X} &= \left[\begin{array}{cc} A_{11} & A_{12} \\ A_{21} & A_{22} \end{array}\right] \,
\left[\begin{array}{cc} X_{11} & X_{12} \\ X_{21} & X_{22} \end{array}\right]
+\left[\begin{array}{cc} X_{11} & X_{12} \\ X_{21} & X_{22} \end{array}\right] \, \left[\begin{array}{cc} A_{11} & A_{12} \\ A_{21} & A_{22} \end{array}\right ]^T \\
	&= \left[\begin{array}{cc} A_{11} & A_{12} \\ A_{21} & A_{22} \end{array}\right] \,
\left[\begin{array}{cc} X_{11} & X_{12} \\ X_{21} & X_{22} \end{array}\right]
+\left[\begin{array}{cc} X_{11} & X_{12} \\ X_{21} & X_{22} \end{array}\right] \,  \left[\begin{array}{cc} A_{11}^T & A_{21}^T \\ A_{12}^T & A_{22}^T \end{array}\right]
\end{align*}
Then, $\dot{X}$ assumes the following formulation.
\begin{align}
	\dot{X} =  \left [ \begin{array}{cc} A_{11} X_{11} +  A_{12} X_{21} + X_{11} A_{11}^T + X_{12} A_{12}^T &
		A_{11} X_{12} + A_{12} X_{22}+ X_{11} A_{21}^T + X_{12} A_{22}^T \\
		\\ A_{21} X_{11} + A_{22} X_{21} + X_{21} A_{11}^T + X_{22} A_{12}^T & A_{21} X_{12} + A_{22} X_{22} + X_{21} A_{21}^T + X_{22} A_{22}^T   \end{array} \right ]
	\label{XdotMatrix}
\end{align}
Recalling that $X_{21}=-X_{12}^T$, we may remark that:
\[  \dot{X}_{11} =  A_{11} X_{11} +   X_{11} A_{11}^T + X_{12} A_{12}^T - A_{12} X_{12}^T \]
\[ \dot{X}_{22}  = A_{22} X_{22} + X_{22} A_{22}^T  + A_{21} X_{12} -  X_{12}^T A_{21}^T   \]
\[ \dot{X}_{12} = A_{11} X_{12} + A_{12} X_{22}+ X_{11} A_{21}^T + X_{12} A_{22}^T \]
Taking $\textrm{vec}(\cdot)$ in both sides of the last equation and exploiting the row vectorisation identity 
$\textrm{vec}(AXB^T) = (A \otimes B) \textrm{vec}(X)$ yields:
\begin{align*} 
\textrm{vec} ( \dot{X}_{12} )  &= \textrm{vec} ( A_{11} X_{12} ) + \textrm{vec} (X_{12} A_{22}^T ) 
+ \textrm{vec} ( A_{12} X_{22} ) + \textrm{vec} ( X_{11} A_{21}^T )
\\ &= (A_{11} \otimes I ) \textrm{vec}(X_{12}) + (I \otimes A_{22} ) \textrm{vec} ( X_{12} ) + (A_{12} \otimes I ) \textrm{vec} ( X_{22} )  + ( I \otimes A_{21} ) \textrm{vec} (X_{11})  
\\&= (A_{11} \oplus A_{22} )\textrm{vec} ( X_{12} ) + ( A_{12} \otimes I ) \textrm{vec} ( X_{22} ) + ( I \otimes A_{21} ) \textrm{vec} ( X_{11} ) 
\\&= (A_{11} \oplus A_{22} )\textrm{vec} ( X_{12} ) + ( A_{12} \otimes I ) M_{n_2} \vec{X}_{22} + ( I \otimes A_{21} ) M_{n_1} 
\vec{X}_{11}.
\end{align*} 
Next, taking the $\vec{ ( \cdot ) }$ operator in both sides of $\dot{X}_{11}$ and $\dot{X}_{22}$ equations yields:
\begin{align*} \dot { \vec{ X}}_{11} &=  \overrightarrow{( A_{11} X_{11} + X_{11} A_{11}^T )} + \overrightarrow{ (X_{12} A_{12}^T - A_{12} X_{12}^T) } \\& =  A_{11}^{[2]} \vec{X}_{11} + L_{n_1} \textrm{vec} (X_{12} A_{12}^T - A_{12} X_{12}^T)
\\& = A_{11}^{[2]} \vec{X}_{11} + L_{n_1} \textrm{vec} (X_{12} A_{12}^T) - L_{n_1} \textrm{vec} (A_{12} X_{12}^T )
\\&= A_{11}^{[2]} \vec{X}_{11} + L_{n_1} ( I_{n_1} \otimes A_{12} ) \textrm{vec} ( X_{12} ) -L_{n_1} ( A_{12} \otimes I_{n_1} ) \textrm{vec}
(X_{12}^T).
\end{align*}
We next make use of matrix $H_{n_1,n_2}$ which converts row vectorisation to column vectorisation, viz. $\textrm{vec} ( X_{12}^T ) = H_{n_1,n_2} \textrm{vec} (X_{12} )$.
Exploiting the latter identity in the previous equation we prove that:
\begin{align*}
\dot { \vec{ X}}_{11} &= A_{11}^{[2]} \vec{X}_{11} +L_{n_1} ( I_{n_1} \otimes A_{12} ) -L_{n_1} ( A_{12} \otimes I_{n_1} ) H_{n_1,n_2} \textrm{vec} (X_{12}).
\end{align*}
Hence, $B_1=  L_{n_1} ( I_{n_1} \otimes A_{12} ) -L_{n_1} ( A_{12} \otimes I_{n_1} ) H_{n_1,n_2}$.
A similar expression can be proved for $\dot{\vec{X}}_{22}$.

\section{Small-gain theorems for stability of $A^{[2]}$}
\label{linear}
We aim to formulate a modular criterion for asymptotic stability of $A^{[2]}$ on the basis of subsystems $\vec{X}_{11}$,
$\vec{X}_{22}$ and $\textrm{vec} (X_{12})$.
To this end we introduce the following notions of gain.
\begin{defn}
\label{def:L2 gains}
For a system of equations:
\[ \dot{x} = A x + B w \]
we denote its $\mathcal{L}_2$ gain as the minimum $\gamma \ge 0$ such that the following LMI admits a positive definite solution $P>0$:
\[  \left [ \begin{array}{cc}  A^T P + P A + I& PB \\ B^T P & -\gamma^2 I   \end{array} \right ] \leq 0 ~.   \]
\end{defn}
Observe that the minimization problem is well posed, despite its open domain.

Consider the $\mathcal{L}_2$ gains $\gamma_1$, $\gamma_2$ and $\gamma_{12}$ for the $\vec{X}_{11}$,
$\vec{X}_{22}$ and $\textrm{vec} (X_{12})$ subsystems of~(\ref{coupledsystem}).
They fuflill the following LMIs:
\begin{align}
	\label{gamma1}
	&\left [ \begin{array}{cc}  {A_{11}^{[2]}}^T P_1 + P_1 A_{11}^{[2]} + I & P_1 B_1 \\ B_1^T P_1 & -\gamma_1^2 I   \end{array} \right ] \leq 0\\ 
	\label{gamma2}
	&\left [ \begin{array}{cc}  {A_{22}^{[2]}}^T P_2 + P_2 A_{22}^{[2]} + I & P_2 B_2 \\ B_2^T P_2 & -\gamma_2^2 I   \end{array} \right ] \leq 0 \\  
	\label{gamma12}
	&\left [\begin{array}{cc}  Q_{12} & P_{12} [G_1, G_2] \\
		{[} G_1, G_2 ]^T P_{12} & -\gamma_{12}^2 I   \end{array} \right ] \leq 0 
\end{align}
where $Q_{12}$ is defined as
\begin{align}
\label{eq:Q12}
	  &Q_{12} = {(A_{11} \oplus A_{22})}^T P_{12} + P_{12} (A_{11} \oplus A_{22}) + I ~.
\end{align}

In the case of subsystem  $\textrm{vec} (X_{12})$ in (\ref{coupledsystem}),  an alternative notion of $\mathcal{L}_2$ gain can be introduced when the input vector is explicitly partitioned into two different signals.
\begin{defn}
\label{def:eta1eta2}
For a linear system:
	\[ \dot{x} = Ax + B_1 u_1 + B_2 u_2 \]
whose input vector is decomposed according to $u=[u_1^T, u_2^T]^T$, the ``partitioned'' $\mathcal{L}_2$ gains $\eta_1 \ge 0$ and $\eta_2 \ge 0$ are defined with respect to the individual inputs components, provided that the following condition
	\[
	\left [ \begin{array}{ccc}  A^T P + P A + I & P B_1 & P B_2 \\ B_1^T P & - \eta_1^2 I & 0 \\ B_2^T P & 0 & - \eta_2^2 I    \end{array}\right ] \leq 0
	\]
	holds for some symmetric matrix $P >0$.
\end{defn}

Definition~\ref{def:eta1eta2} implies that the partitioned $\mathcal{L}_2$ gains $\eta_{1}$ and $\eta_2$ of subsystem $\textrm{vec} (X_{12})$ in~(\ref{coupledsystem}) fulfill for some $P_{12}=P_{12}^T>0$ the following LMI:
\begin{equation}
	\label{eq:gamma12 LMI}
	\left [ \begin{array}{ccc}  Q_{12} & P_{12} G_1 & P_{12} G_2 \\ G_1^T P_{12} & - \eta_1^2 I & 0 \\ G_2^T P_{12} & 0 & - \eta_2^2 I    \end{array}\right ] \leq 0
\end{equation}
where $Q_{12}$ is as in (\ref{eq:Q12}).
\begin{rem}
It is worth noting that in Definition~\ref{def:eta1eta2}, differently from Definition~\ref{def:L2 gains}, we do not look for the minimal values $\eta_1$, $\eta_2$ of such ``partitioned'' gains, as a priori it is not obvious that one can simultaneously minimize $\eta_1$ and $\eta_2$.
In particular, a non-trivial Pareto front of minimal values for $\eta_1$ and $\eta_2$ might occur, i.e., they may not be independent from each other.
\end{rem}

Our main results are the following small-gain theorems.
\begin{thm}
\label{firstmain}
Consider an interconnected system formulated as in~(\ref{interconnected}), and let~$\gamma_1$ and~$\gamma_2$ denote the~$\mathcal{L}_2$ gains of subsystems~$A_{11}^{[2]}$ and~$A_{22}^{[2]}$ computed according to Definition~\ref{def:L2 gains} and (\ref{gamma1})-(\ref{gamma2}). 
Then, the related second additive compound matrix~$A^{[2]}$ is Hurwitz, if $\exists\, \varepsilon > 0$ such that the following small-gain condition is satisfied
\begin{align}
\label{relaxedsmallg}
	& 1 > \min_{ P_{12}, \tilde{\eta}_1, \tilde{\eta}_2 } \gamma_1^2 \tilde{\eta}_1 + \gamma_2^2 \tilde{\eta}_2 \\
	& \left[\begin{array}{ccc}
		Q_{12} & P_{12} G_1 & P_{12} G_2 \\
		G_1^T P_{12} & - \tilde{\eta}_1 I & 0 \\
		G_2^T P_{12} & 0 & - \tilde{\eta}_2 I
	  \end{array}\right ] \leq 0 \notag\\
	& P_{12}=P_{12}^T \geq \varepsilon I \notag\\
	& \tilde{\eta}_1 \geq 0,  \quad \tilde{\eta}_2 \geq 0 \notag
\end{align}
where $Q_{12}$ is as in (\ref{eq:Q12}).
\end{thm}
\emph{Proof.} - See the Appendix.

A similar result can be formulated without involving the partitioned $\mathcal{L}_2$ gains of Definition~\ref{def:eta1eta2}, as follows.
\begin{thm}
\label{secondmain}
Consider an interconnected system formulated as in~(\ref{interconnected}).
The related second additive compound matrix~$A^{[2]}$ is Hurwitz if the~$\mathcal{L}_2$ gains~$\gamma_1$, $\gamma_2$ and~$\gamma_{12}$, computed computed according to Definition~\ref{def:L2 gains} and (\ref{gamma1})-(\ref{gamma12}), fulfill the small-gain condition:
\begin{equation}
\label{sgt2}
    \gamma_{12} \cdot \sqrt{ \gamma_1^2 + \gamma_2^2 }  < 1.
\end{equation}
\end{thm}
\emph{Proof.} - See the Appendix.
\begin{rem} It is interesting to remark that $A_{11}^{[2]}$, $A_{22}^{[2]}$ and $A_{11} \oplus A_{22}$ may be Hurwitz matrices even if $A_{11}$ or $A_{22}$ are not. In particular, asymptotic stability of the individual subsystems is not a necessary condition for the application of the proposed small-gain conditions to the stability of $A^{[2]}$.
\end{rem}
\begin{rem}
	\label{rem:cascade}
	In the case of systems in cascade, i. e. when $A$ is block-triangular, the traditional small-gain condition for asymptotic stability of interconnected systems is automatically satisfied, as soon as one of the gains of the two subsystems is zero, because the loop gain becomes~$\gamma_1 \gamma_2 = 0 <1$.
	One would expect something similar to hold in small-gain conditions for $2$-contraction.
	In this regard, observe that when matrix~$A$ is block-triangular, so is~$A^{[2]}$, albeit up to permutation of variables.
	For instance, $A_{12}=0$ implies $B_1=0$ and $G_2=0$ in equation~(\ref{coupledsystem}).
	As a result, the evolution of $X_{11}$ describes an autonomous system, which feeds into subsystem $X_{12}$, that in turn forces subsystem $X_{22}$.
	Therefore, if matrices $A_{11}^{[2]}$, $A_{11} \oplus A_{22}$ and $A_{22}^{[2]}$ are Hurwitz, so is matrix $A^{[2]}$ and the overall system is $2$-contracting.
	Then, consider the small-gain condition~(\ref{relaxedsmallg}) of Theorem~\ref{firstmain} and let us denote by $P_{12}^*$, $\eta_1^*$ and $\eta_2^*$ the optimal values of $P_{12}$, $\eta_1$ and $\eta_2$, respectively.
	In the case of cascaded systems this condition is tight, because $B_1=0$ and $G_2=0$ allows for computing $\gamma_1=0$ and selecting $\eta_2^*=0$, so that $\gamma_1^2 \eta_1^* + \gamma_2^2 \eta_2^* = 0 < 1$ is guaranteed.
	Conversely, in Theorem~\ref{secondmain}, where a unique gain~$\gamma_{12}$ is adopted to characterize the amplification introduced by subsystem $X_{12}$, the small-gain condition~(\ref{sgt2}) is not automatically fulfilled.
	In the case exemplified, for instance, $\gamma_{1}=0$, but this still requires the fulfillment of the condition $\gamma_{12} \cdot \gamma_2 <1$ to guarantee $2$-contraction of the cascade according to Theorem~\ref{secondmain}.
	This situation is unideal as it hints at some conservatism in this latter formulation.
\end{rem}

\section{Modular 2-contraction of nonlinear systems}
We consider next the case of interconnected nonlinear systems, defined by $\mathcal{C}^1$ equations:
\begin{equation}
\label{interconnected2}
 \left [ \begin{array}{c}\dot{x}_1 \\ \dot{x}_2 \end{array} \right ]
 = \left [ \begin{array}{c} f_1 (x_1,x_2) \\ f_2 (x_1, x_2) \end{array} \right ] = : f(x), 
\end{equation}
where $x_1$ and $x_2$ are vectors of dimension $n_1, n_2 \geq 2$. Due to the smoothness of $f_1$ and $f_2$ we may define the block-partitioned Jacobian matrix $J$ given below:
\begin{equation}
    \label{jacobianpartition}
    J(x) = \left[\begin{array}{cc}
    \frac{ \partial f_1}{\partial x_1} (x) & \frac{ \partial f_1}{\partial x_2} (x) \\ \\
    \frac{ \partial f_2}{\partial x_1} (x) & \frac{ \partial f_2}{\partial x_2} (x) 
    \end{array}\right]~.
\end{equation}
It was shown in~\cite{muldowney} that suitable contraction conditions (expressed through matrix norms) of the second additive compound of the Jacobian $J^{[2]}(x)$, can be used to rule out periodic solutions in nonlinear dynamical systems.
Such conditions were reformulated in~\cite{muhammadIEEE,IEEEmartini} through the use of Lyapunov functions or LMIs and extended to rule out oscillatory behaviors of periodic, almost periodic and chaotic nature.
The goal of this section is to exploit/extend the modular criteria proposed in Section~\ref{linear} to the case of interconnected nonlinear systems as given by (\ref{interconnected2}) in order to rule out oscillatory behaviors.

We work under the assumption that a compact forward invariant set $\mathcal{X} \subseteq \mathbb{R}^{n}$ for the dynamics
of (\ref{interconnected2}) is available or that solutions are a priori known to be bounded. 
Then, oscillatory behaviors may be ruled out provided 
a symmetric and positive definite $x$-dependent matrix
$P(x) \in \mathbb{R}^{n \choose 2}$ is known to satisfy both $\alpha_1 I \leq P(x) \leq \alpha_2 I$, for positive $\alpha_1,\alpha_2$, and
\begin{equation}
\label{contractionLMI}
    J^{[2]}(x)^{T} P(x) + P(x) J^{[2]} (x) + \dot{P} (x) \leq - \varepsilon I
\end{equation}
for some $\varepsilon>0$ and $\forall x \in \mathcal{X}$. \\
Similarly to the linear case, the variational equation associated to the second-order additive compound matrix, i.e.
\begin{equation}
\label{standardvar}
\begin{array}{rcl}
\dot{x} & = & f(x) \\
\dot{ \delta}^{[2]} & = & J^{[2]} (x) \, \delta^{[2]}    
\end{array}
\end{equation}
can be rearranged according to equation (\ref{coupledsystem}) as
\begin{equation}
\label{modularvariational}
\left\{
	\begin{array}{l}
	\dot{x} = f(x) \\
	\dot{\delta}_1 = J_{11}^{[2]} (x) \, \delta_1 + B_1 (x) \delta_{12} \\
	\dot{ \delta}_{12} = ( J_{11} (x) \oplus J_{22} (x) ) \delta_{12} + G_1(x) \delta_1 + G_2(x) \delta_2 \\
	\dot{\delta}_2 = J_{22}^{[2]} (x) \, \delta_2 + B_2 (x) \delta_{12}.
	\end{array}\right.
\end{equation}
Condition (\ref{contractionLMI}) ensures exponential convergence of $\delta^{[2]}(t)$ for any initial condition in $\mathcal{X}$ and
any initial value of $\delta^{[2]} (0) \in \mathbb{R}^{n \choose 2}$.
Our goal is the formulation of small-gain conditions analogous to (\ref{relaxedsmallg}) and (\ref{sgt2}) to ensure (\ref{contractionLMI}).
To this end we introduce the notion of $\mathcal{L}_2$ gain
for state dependent matrices according to the following LMIs.
\begin{defn}
\label{l2 gain x}
	For a system of equations:
	\[ \dot{\delta} = A(x) \delta + B(x) w \]
	we define its $\mathcal{L}_2$ gain as any value $\gamma \ge0$ such that the LMI:
	\[  \left [ \begin{array}{cc}  A(x)^T P(x) + P(x) A(x) + \dot{P}(x) + I & P(x) B(x) \\ B(x)^T P(x) & -\gamma^2 I   \end{array} \right ] \leq 0.   \]
	is fulfilled for all $x \in \mathcal{X}$ and for some positive definite symmetric matrix function $P(x)$ of class $\mathcal{C}^1$.
\end{defn}
It is worth pointing out that $\dot{P}(x)$ is the matrix of entries $[ L_f P_{ij} (x)]$ with $i,j \in 1, \ldots n$ and $L_f$ denotes the Lie derivative along solutions of $\dot{x} = f(x)$.

We can now define the gains of the $\delta_1$, $\delta_2$ and $\delta_{12}$ subsystems in (\ref{modularvariational}). In particular, we say that $\gamma_1$ is the gain of the $\delta_1$ subsystem if for some $P_1(x)$ of class $\mathcal{C}^1$ and all $x \in \mathcal{X}$ it fulfills
\begin{equation}
	\label{j1gain}
	\left [ \begin{array}{cc}  Q_1(x)
		& P_1(x) B_1(x)  \\  B_1(x)^T P_1(x) & - \gamma_1^2 I \end{array} \right ] \leq 0 \\
\end{equation}
where
\begin{equation*}
Q_1(x) = J_{11}^{[2]} (x) P_1(x) + P_1(x) J_{11}^{[2]} (x) + \dot{P}_1(x) + I ~.	
\end{equation*}
Similarly, for subsystem $\delta_2$ the gain $\gamma_2$ is computed according to the following LMI condition:
\begin{equation}
	 \label{j2gain}
	 \left [ \begin{array}{cc}  Q_2(x)
	 	& P_2(x) B_2(x)  \\  B_2(x)^T P_2(x) & - \gamma_2^2 I \end{array} \right ] \leq 0\\
\end{equation}
where
\begin{equation*}
	Q_2(x)=J_{22}^{[2]} (x) P_2(x) + P_2(x) J_{22}^{[2]} (x) + \dot{P}_2(x) + I ~.
\end{equation*}
Finally, the gain $\gamma_{12}$ of the component $\delta_{12}$ of the variational equation (\ref{modularvariational}) is given by the fulfillment of
\begin{equation}
\label{j12gain}
	\left[\begin{array}{cc}
	Q_{12}(x) & P_{12}(x) {[} G_1(x), G_2(x) {]}   \\
	{[} G_1(x), G_2(x) {]}^T P_{12}(x) & -\gamma_{12}^2 I
	\end{array}\right] \leq 0
\end{equation}
where 
\begin{align*}
	Q_{12}(x) =& ( J_{11}(x) \oplus J_{22}(x)  )^T P_{12}(x) 
	+ P_{12}(x) ( J_{11}(x) \oplus J_{22}(x)) +\dot{P}_{12}(x) + I ~.
\end{align*}
As for the Definition~\ref{def:eta1eta2}, when a system has the same formulation of subsystem $\delta_{12}$ in (\ref{modularvariational}), an alternative notion of $\mathcal{L}_2$ gain can be introduced.
\begin{defn}
\label{def:definition 4}
For a system of equations:
\[ \dot{\delta} = A(x) \delta + B_1(x) u_1 + B_2(x) u_2  \]
whose input vector is decomposed according to $u=[u_1^T, u_2^T]^T$, the ``partitioned'' $\mathcal{L}_2$ gains $\eta_1$ and $\eta_2$ are defined with respect to those individual input components, provided that the LMI condition:
\begin{align*}  
\left[ \begin{array}{ccc} 
	Q(x) & P(x) B_1(x) & P(x) B_2(x) \\
	B_1(x)^T P(x) & -\eta_1^2 I & 0 \\
	B_2(x)^T P(x) & 0 & -\eta_2^2 I
\end{array}\right] \leq 0 ~,
\end{align*}
where
\begin{equation*}
	Q(x) = A(x)^T P(x) + P(x) A(x) + I +\dot{P}(x) ~,
\end{equation*}
holds for some positive definite symmetric matrix function $P(x)$ of class $\mathcal{C}^1$ and for all the $x \in \mathcal{X}$.
\end{defn}
%

Definition~\ref{def:definition 4} implies that the partitioned $\mathcal{L}_2$ gains $\eta_{1}$ and $\eta_2$ of subsystem $\delta_{12}$ in~(\ref{modularvariational}) fulfill the following LMI 
\begin{equation}
	\label{eta1eta2nl}
	\left[ \begin{array}{ccc}
		Q_{12}(x) & P_{12}(x) G_1(x) & P_{12}(x) G_2(x) \\
		G_1^T(x) P_{12}(x) & - \eta_1^2 I & 0 \\
		G_2^T(x) P_{12}(x) & 0 & - \eta_2^2 I
	\end{array}\right] \leq 0 ,
\end{equation}
where 
\begin{align}
	Q_{12}(x) =& (J_{11}(x) \oplus J_{22}(x))^T P_{12}(x)
	+P_{12}(x) (J_{11}(x) \oplus J_{22}(x))  +\dot{P_{12}}(x) + I ,
	\label{eq:Q bar}
\end{align}
for some positive definite matrix function $P_{12}(x)$ of class $\mathcal{C}^1$ and for all $x \in \mathcal{X}$.
\begin{rem}
While it is in principle possible to use state-dependent matrices $P_1(x)$, $P_2(x)$ and $P_{12}(x)$ for the definition of the gains,
as in~(\ref{j1gain}), (\ref{j2gain}), (\ref{j12gain}), and~(\ref{eta1eta2nl}), the computation of the derivatives $\dot{P}_1(x)$, $\dot{P}_{12}(x)$, and $\dot{P}_2(x)$ cannot be done in a decoupled fashion. In this respect, a noteworthy simplification occurs when dealing with constant matrices, as the gains can be computed independently of each other. Namely, changing $f_2(x_1,x_2)$ for $\dot{x}_2$ will not affect the gain $\gamma_1$ of the $\delta_1$ subsystem and vice-versa. On the other hand the $\gamma_{12}$ gain is affected both by $\dot{x}_1$ and $\dot{x}_2$.
An intermediate situation can be pursued by choosing $P_1(x_1)$, $P_2 (x_2)$ and $P_{12}$ constant, so as to still retain some decoupling in the computation of gains and allow the flexibility of state-dependent matrices.
\end{rem}
\begin{thm}
\label{nonlinearsgt1}
Consider the interconnected system (\ref{interconnected2}) and assume that everywhere in some forward invariant set its Jacobian matrix~(\ref{jacobianpartition}) belongs to the convex hull of the set of matrices $\mathcal{V}=\left\{V_i\right\}_{i=1,\ldots,m}$, i.e., $J(x)\in \textrm{conv}(\mathcal{V})$ for all $x \in \mathcal{X}$.
Let~$\gamma_1$ and~$\gamma_2$ be computed according to Definition~\ref{l2 gain x} and~(\ref{j1gain})-(\ref{j2gain}). 
	Then, the second additive compound matrix of the Jacobian $J^{[2]} (x)$ fulfills the contraction property~(\ref{contractionLMI}), if $\exists\, \varepsilon > 0$ such that the following small-gain condition is satisfied for all the matrices $V_i$
\begin{align}
\label{relaxedsmallgNL}
	& 1 >  \min_{ P_{12}, \tilde{\eta_1}, \tilde{\eta_2} } \gamma_1^2 \tilde{\eta_1} + \gamma_2^2 \tilde{\eta_2} \\
\label{hull_relaxed_cond}
	& \left[\begin{array}{ccc}
		Q_{12,i}(x) & P_{12} G_{1,i} & P_{12} G_{2,i} \\
		G_{1,i}^T P_{12} & -\tilde{\eta}_1 I & 0 \\
		G_{2,i}^T P_{12} & 0 & -\tilde{\eta}_2 I
	  \end{array}\right ] \leq 0 ~,~ i=1,\ldots,m  \\
	& P_{12} = P_{12}^T \geq \varepsilon I \notag\\
	& \tilde{\eta_1} \geq 0, ~ \tilde{\eta_2} \geq 0 \notag
\end{align}
where $Q_{12,i}(x)$ has the same form as in~(\ref{eq:Q bar}) but $P_{12}$ does not depend on $x$ and $J(x)$ is played by $V_i$.
\end{thm}
\emph{Proof.} - See the Appendix.

\begin{thm}
\label{nonlinearsgt2}
Consider the interconnected system (\ref{interconnected2}). The second additive compound matrix of its Jacobian $J^{[2]} (x)$ fulfills the contraction property (\ref{contractionLMI}) if the~$\mathcal{L}_2$ gains~$\gamma_1$, $\gamma_2$ and~$\gamma_{12}$, computed according to Definition~\ref{l2 gain x} and (\ref{j1gain})-(\ref{j12gain}), satisfy the small-gain condition:
\begin{equation}
\label{sgt3}
    \gamma_{12} \cdot \sqrt{ \gamma_1^2 + \gamma_2^2 }  < 1.
\end{equation}
\end{thm}
\emph{Proof.} - See the Appendix.

\begin{rem}
\label{Rem_n3}
It is worth noting that in Section~\ref{sec:prelim} the dimension $n_1$ and $n_2$ of the subsystems' states are limited to be greater or equal than two.
However, the approach can be applied in the case of systems of dimension $n=3$ as well.
In such case, one of the two subsystems is empty and one is scalar, for which the gain $\gamma_1$ can be readily computed.
Therefore, conditions~(\ref{sgt2}) and~(\ref{sgt3}) become:
\begin{equation}
    \gamma_{12} \cdot \gamma_1 < 1.
\end{equation}
\end{rem}

\section{Examples of application}
\subsection{Transition from multistability to limit cycles}
\label{sec:example 1}
\begin{figure}[tb]
    \centering
    \includegraphics[width=0.66\columnwidth]{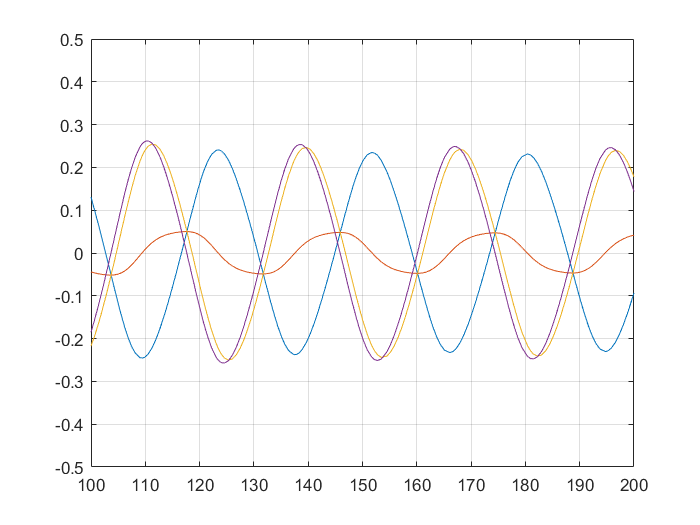}
    \caption{Oscillatory solution of system~(\ref{ex:Angeli system}), when the configuration is set at $k=1.1$, a value for which the model can be regarded as the feedback interconnection of the two subsystems illustrated in Section~\ref{sec:example 1}.}
    \label{oscillatoryfig}
\end{figure}
Consider the system of equations:
\begin{equation}
    \begin{array}{rcl}
	\dot{x}_1 & = & x_2 \\
	\dot{x}_2 & = & - x_1 + \textrm{atan}(2 x_1 ) -2 x_2 + x_3 \\
	\dot{x}_3 & = & -x_3 + x_4 \\
	\dot{x}_4 & = & - k x_1 - x_ 4
	\end{array}~.
	\label{ex:Angeli system}
\end{equation}
The system can be regarded as the feedback interconnection of the $(x_1,x_2)$ and $(x_3,x_4)$ subsystems through the linking signals $x_1$ and $x_3$.
Notably, for $k=0$ the system boils down to the cascade (series) interconnection of the asymptotically stable linear subsystem $(x_3,x_4)$, forced with vanishing intensity by the multistable bidimensional subsystem $(x_1,x_2)$.
In this latter case, nonoscillatory behaviors of the multistable system for $x_3 (t) \equiv 0$ can be shown by considering the Lyapunov functional
\[ V(x_1,x_2) = \frac{ x_2^2 }{2} + \int_0^{x_1} \xi - \textrm{atan} ( 2 \xi ) d \xi. \]
Hence, for $k=0$ this system is multistable and it has two asymptotically stable equilibria, and a third, unstable, saddle in~$0$.
Our goal is to find sufficient conditions that guarantee non-oscillatory behaviors ($2$-contraction) of the system also for some range of $k>0$.

It is easy to see, through simulations, that for $k$ sufficiently large the system admits oscillatory solutions, as shown in Fig.~\ref{oscillatoryfig}. In fact, this occurs for all $k>1$.
The Jacobian $J(x)$ is given as:
\[  J(x) = \left [  \begin{array}{cccc} 0 & 1 & 0 & 0 \\
-1 + \frac{2}{1+4 x_1^2} & -2 & 1 & 0 \\ 0 & 0 & -1 & 1 \\ - k & 0 & 0 & -1
\end{array}
\right ]\]
Notice that, no matter what $x_1$ is, the Jacobian $J(x)$ belongs to the interval matrix
\[
J(x) \in \left[\begin{array}{cccc} 0 & 1 & 0 & 0 \\
{[}-1,1] & -2 & 1 & 0 \\ 0 & 0 & -1 & 1 \\ - k & 0 & 0 & -1
\end{array}\right]
=\textrm{conv}\left(J_1(k),J_2(k)\right)
\]
where $J_1(k)$ and $J_2(k)$ read
\[\begin{split} &J_1(k) = \left [  \begin{array}{cccc} 0 & 1 & 0 & 0 \\
-1 & -2 & 1 & 0 \\ 0 & 0 & -1 & 1 \\ - k & 0 & 0 & -1
\end{array}
\right ], \\ 
&J_2(k) =\left [  \begin{array}{cccc} 0 & 1 & 0 & 0 \\
1 & -2 & 1 & 0 \\ 0 & 0 & -1 & 1 \\ - k & 0 & 0 & -1
\end{array}
\right ].\end{split}\]
Rather than considering the full $J(x)$, and the corresponding $J^{[2]}$ matrix (of dimension $6 \times 6$), we decompose the system into its $(x_1,x_2)$ and $(x_3,x_4)$ components, respectively.
Notice that standard small-gain results do not apply, as $J(0)$, even for $k=0$, has a positive eigenvalue in $\frac{ -1+\sqrt{2} }{2}$.
The modular version of the second additive compound variational equation looks like:
\begin{equation}
 \begin{split}
&\dot{\delta}_1  =  -2  \delta_1 + [1,0,0,0] \, \delta_{12} \\ 
&\begin{split}\dot{\delta}_{12} &= \left [ \begin{array}{cccc} -1 & 1 & 1 & 0 \\ 0 & -1 & 0 & 1 \\ -1 + \frac{2}{1+(2x_1)^2} & 0 & -3 & 1 \\
0 & -1 + \frac{2}{1+(2x_1)^2} & 0 & -3 \end{array} \right ] \delta_{12} + \left [\begin{array}{c} 0 \\  0 \\  0 \\ k  
\end{array} \right ] \, \delta_1 + \left [\begin{array}{c} 0 \\  0 \\  0 \\ 1  
\end{array} \right ] \, \delta_2 \end{split}\\
&\dot{\delta}_2  =  - 2 \delta_2 + [k,0,0,0] \delta_{12}
    \end{split}
\end{equation}
It is easy to see that $\gamma_1=1/2$ and $\gamma_2=k/2$.
To apply Theorem~\ref{nonlinearsgt1} we need to solve the following minimization problem:
\begin{align}
	\label{SG_relaxedLMI1}
	& \min_{ P_{12}, \tilde{\eta}_1, \tilde{\eta}_2 } \gamma_1^2 \tilde{\eta}_1 + \gamma_2^2 \tilde{\eta}_2   < 1 \\
	& \textrm{subject to} \notag\\
	& \left[\begin{array}{ccc}
		A_1^T P_{12} + P_{12}  A_1 + I & P_{12} G_1 & P_{12} G_2 \\
		G_1^T P_{12} & - \tilde{\eta}_1 I & 0 \\
		G_2^T P_{12} & 0 & - \tilde{\eta}_2 I
	  \end{array}\right ] \leq 0  \notag\\
	& \left [ \begin{array}{ccc}
		A_2^T P_{12} + P_{12}  A_2 + I & P_{12} G_1 & P_{12} G_2 \\
		G_1^T P_{12} & - \tilde{\eta}_1 I & 0
		\\ G_2^T P_{12} & 0 & - \tilde{\eta}_2 I
	  \end{array}\right ] \leq 0  \notag\\
	& P_{12} = P_{12}^T \geq \varepsilon I  \notag\\
	& \tilde{\eta}_1 \geq 0  \notag\\
	& \tilde{\eta}_2 \geq 0  \notag
\end{align} 
where the matrices $A_1$ and $A_2$ are given by
\[\begin{split} & A_1 = \left [ \begin{array}{cccc} -1 & 1 & 1 & 0 \\ 0 & -1 & 0 & 1 \\ -1  & 0 & -2 & 1 \\
		0 & -1  & 0 & -2 \end{array} \right ], \\ &A_2 = \left [ \begin{array}{cccc} -1 & 1 & 1 & 0 \\ 0 & -1 & 0 & 1 \\ 1  & 0 & -2 & 1 \\
		0 & 1  & 0 & -2 \end{array} \right ] , \end{split}\]
and $\varepsilon=0.01$. The maximum value of the parameter $k$ for which the minimization problem~(\ref{SG_relaxedLMI1}) turns out to be feasible is $k^* = 0.755$, and it is obtained for
\[
	P_{12} = \left[\begin{array}{cccc}
	1.24  &  0.68  &  0.14  &  0.19\\
	0.68  &  4.98  &  0.18  &  1.48\\
	0.14  &  0.18  &  0.85  &  0.19\\
	0.19  &   1.48 &   0.19 &   1.45
	\end{array}\right]
\]

\begin{figure}[tb]
	\centering
	\includegraphics[width=0.66\columnwidth]{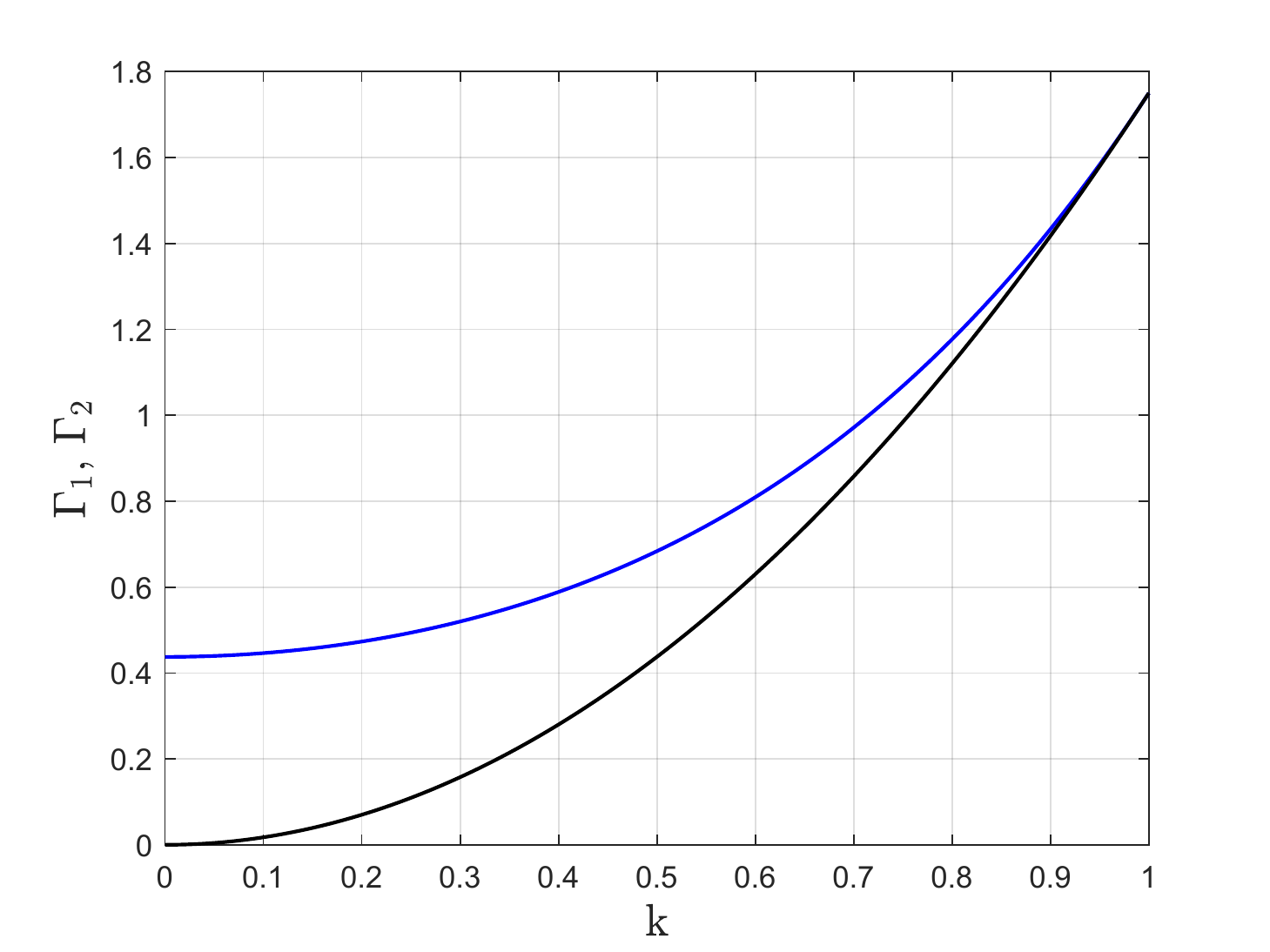}
	\caption{Functions $\Gamma_1(k)\doteq\gamma_{1}^2(k)\tilde{\eta}_1(k)+\gamma_2^2(k)\tilde{\eta_2}(k)$ (black) and $\Gamma_2(k)\doteq\gamma_{12}^2(k)(\gamma_{1}^2(k)+\gamma_2^2(k)$ (blue) as a function of the parameter $k\in[0,1]$.
		\label{Cond_ASys}}
\end{figure}

Instead, exploiting Theorem~\ref{nonlinearsgt2}, the maximum gain $\gamma_{12}(k)$ allowed by the small-gain condition (\ref{sgt3}) as a function of parameter $k$ is given by $\gamma_{12}(k) = 1/ \sqrt{(1/2)^2+(k/2)^2}$. 
In this case, we have to solve the following maximization problem:
\begin{align} 
\label{SG_lmi1}
	& \max_{ k \geq 0, P_{12} = P_{12}^T }  k \\
	& \textrm{subject to} \notag\\
	& \left[\begin{array}{cc}
		A_1^T P_{12} +P_{12} A_1 + I & P_{12} [ G_1, G_2 ] \\
		{[} G_1 G_2 ]^T P_{12} & - \gamma_{12}^2(k) I
	  \end{array} \right ] \leq 0  \notag\\
	& \left[\begin{array}{cc}
		A_2^T P_{12} +P_{12} A_2 + I & P_{12} [ G_1, G_2 ] \\
		{[} G_1 G_2 ]^T P_{12} & - \gamma_{12}^2(k) I
	  \end{array}\right] \leq 0  \notag\\
	& P_{12} \geq  0 \notag
\end{align}
The maximum value of the parameter $k$ for which the maximization problem (\ref{SG_lmi1}) is feasible is $k^* = 0.715$, which is obtained for
\[
P_{12} = \left[\begin{array}{cccc}
     1.25 &   0.68 &   0.13  &  0.18\\
    0.68  &  5.0  &  0.17   & 1.47\\
    0.13  &  0.17  &  0.89  &  0.19\\
    0.18  &  1.47  &  0.19  &  1.49
\end{array}\right]~.
\]
In Fig.~\ref{Cond_ASys} are reported the conditions (\ref{relaxedsmallgNL}) and (\ref{sgt3}) as a function of the parameter $k$.
From the figure it can be observed that the condition in Theorem~\ref{nonlinearsgt2} is more conservative than the condition in Theorem~\ref{nonlinearsgt1}, as introduced in Remark~\ref{rem:cascade}.

To measure the conservativeness of the small-gain conditions (\ref{relaxedsmallgNL}) and (\ref{sgt3}), we compare the value $k^*$ with the one achievable by means of the following maximization problem
\begin{align}
\label{SG_lmi2}
	& \max_{ k \geq 0, P=P^T } k \\
	& \textrm{subject to} \notag\\ 
	& J_1^{[2]}(k)^T P + P J_1^{[2]}(k) \leq 0 \notag\\
	& J_2^{[2]}(k)^T P + P J_2^{[2]}(k) \leq 0 \notag\\
	& P \geq I \notag
\end{align} 
which directly involves the additive compound matrix and a proper matrix $P$ of dimension $6\times6$. It turns out that problem (\ref{SG_lmi2}) is feasible for all $k\in [0,1]$ and that the maximum value is achieved for
\[
P = \left[\begin{array}{cccccc}
   11.45 & 0  &  2.43  & -0.209  &  1.99 &  -1.01\\
   0 &  14.09 &  10.88 &   1.37&    3.51&    0\\
   2.43  & 10.88 &  37.17  &  2.14 &   7.31 &   2.43\\
   -0.209 & 1.37 &   2.14   & 8.49  &  1.70  & -0.21\\
   1.99  & 3.51 &   7.31   & 1.70  &  9.29  &  1.99\\
   -1.01 & 0 &   2.43   &-0.21  &  1.99  & 11.45
\end{array}\right]~.
\]
It is worth nothing that for values of $k$ greater than 1 the system starts to display periodic motions (see Fig.~\ref{oscillatoryfig}).

\subsection{Thomas' example of dimension 4}
\label{sec:Thomas 4}
\begin{figure}[tb]
    \centering
    \includegraphics[width=0.66\columnwidth]{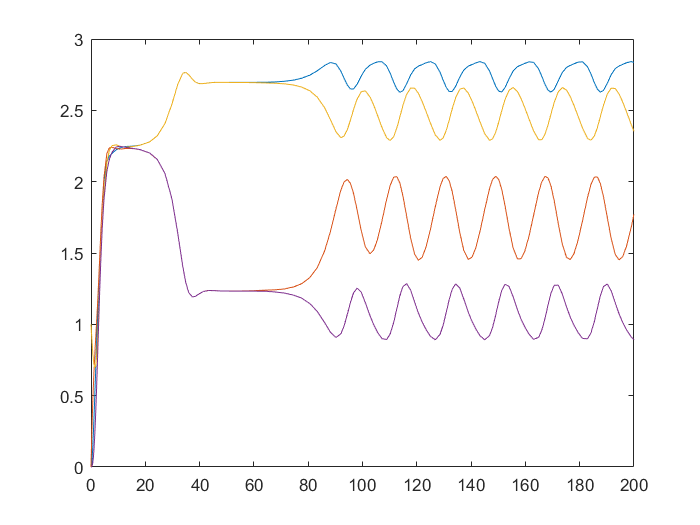}
    \caption{The fourth order Thomas' system~(\ref{ex:Thomas 4}) of Section~\ref{sec:Thomas 4} shows periodic solutions when it is configured with $b=0.35$.}
    \label{oscillatorythomas}
\end{figure}
As a further example, let us consider the Thomas' system (see~\cite{thomas}) of the fourth order, described by the following set of first order differential equations
\begin{equation}
    \begin{array}{rcl}
	\dot{x}_1 & = & - b x_1 + \sin(x_2)\\
	\dot{x}_2 & = & - b x_2 + \sin(x_3) \\
	\dot{x}_3 & = & - b x_3 + \sin(x_4) \\
	\dot{x}_4 & = & - b x_4 + \sin(x_1)
	\end{array} ~,
	\label{ex:Thomas 4}
\end{equation}
where $b$ is a positive scalar parameter. For $b>1$ the system has a unique asymptotically stable equilibrium point at $x=0$, which undergoes a (supercritical) pitchfork bifurcation at $b=1$. For $b<1$ the system is multistable and it exhibits quite a rich dynamic behavior as $b$ is decreased towards $0$. For $b=0.35$, periodic solutions arise, as seen from numerical simulations depicted in Fig.~\ref{oscillatorythomas}.
Moreover, the hypercube $[-1/b,1/b]^4$ is a forward invariant set for system~(\ref{ex:Thomas 4}).\\
Our aim is to find conditions, similar to conditions (\ref{SG_lmi1})-(\ref{SG_lmi2}), to rule out oscillatory behaviors for some range of $b<1$.

The Jacobian $J(x)$ of the system has the following form
\[
    J(x) =\left[\begin{array}{cccc}
         -b & c_2 & 0 & 0\\
         0 & -b & c_3 & 0\\
         0 & 0 & -b & c_4\\
         c_1 & 0 & 0 & -b 
    \end{array}\right] ~,
\]
where $c_i = \cos(x_i)$.
It is worth noting that, since $c_i \in [-1,1]$, then $J(x)\in\textrm{conv}(\mathcal{V})$, where $\mathcal{V}=\left\{V_i\right\}_{i=1,\ldots,16}$ collects the Jacobian matrices at the vertexes of the hypercube, i.e.,
\[
\textrm{conv}(\mathcal{V}) =
	\left[\begin{array}{cccc}
         -b & [-1,1] & 0 & 0\\
         0 & -b & [-1,1] & 0\\
         0 & 0 & -b & [-1,1]\\
         {[}-1,1] & 0 & 0 & -b 
    \end{array}\right] ~.
\]
Its second additive compound reads
\[
J^{[2]} (x)= \left[\begin{array}{cccccc} -2\,b & c_{3} & 0 & 0 & 0 & 0\\ 0 & -2\,b & c_{4} & c_{2} & 0 & 0\\ 0 & 0 & -2\,b & 0 & c_{2} & 0\\ 0 & 0 & 0 & -2\,b & c_{4} & 0\\ -c_{1} & 0 & 0 & 0 & -2\,b & c_{3}\\ 0 & -c_{1} & 0 & 0 & 0 & -2\,b \end{array}\right] ~.
\]
We choose to partition the state-space according to $(x_1,x_3)$ and $(x_2,x_4)$.
Therefore, the modular version of the second additive compound variational equation of the fourth order Thomas' system assumes the following form:
\begin{equation}
    \begin{split}
&\dot{\delta}_1  =  -2 b  \delta_1 + [0,c_4,-c_3,0] \, \delta_{12} \\ 
&\begin{split}\dot{\delta}_{12} &= \left [ \begin{array}{cccc} -2b & 0 & 0 & 0 \\ 0 & -2b & 0 & 0 \\ 0 & 0 & -2b & 0 \\
0 & 0 & 0 & -2b \end{array} \right ] \delta_{12} + \left [\begin{array}{c} c_2 \\  0 \\  0 \\ -c_1  
\end{array} \right ] \, \delta_1 + \left [\begin{array}{c} 0 \\  c_3 \\  -c_4 \\ 0  
\end{array} \right ] \, \delta_2 \end{split}\\
&\dot{\delta}_2  =  - 2 b\delta_2 + [c_1,0,0,c_2] \delta_{12}
    \end{split}
\end{equation}
The gains $\gamma_1$ and $\gamma_2$ can be readily computed, obtaining $\gamma_1 = \gamma_2 = 1/\sqrt{2}b$.
Our aim is to solve the minimization problem 
{\small
\begin{align}
\label{SG_relaxedLMI2}
	& \min_{ P_{12}, \tilde{\eta_1}, \tilde{\eta_2} } \gamma_1^2 \tilde{\eta}_1 + \gamma_2^2 \tilde{\eta}_2   < 1 \\
	& \textrm{subject to} \notag\\
	& \left[\begin{array}{ccc}
		A^T P_{12} + P_{12}  A + I & P_{12} G_1^{(h)} & P_{12} G_2^{(h)} \\
		G_1^{(h)T} P_{12} & - \tilde{\eta}_1 I & 0 \\
		G_2^{(h)T} P_{12} & 0 & -\tilde{\eta}_2 I
	  \end{array}\right] \leq 0
	,\, h=1,\ldots,16 \notag\\
	& P_{12}=P_{12}^T \geq \varepsilon I \notag\\
	& \tilde{\eta}_1 \geq 0 \notag\\
	& \tilde{\eta}_2 \geq 0 \notag
\end{align}}\normalsize
where $\varepsilon=0.01$,
\[
\begin{split}
	&A = \left [ \begin{array}{cccc} -2b & 0 & 0 & 0 \\ 0 & -2b & 0 & 0 \\ 0  & 0 & -2b & 0 \\
		0 & 0  & 0 & -2b \end{array} \right ], \\&   G^{(h)} = \left [ \begin{array}{cc} v_2^{(h)} & 0  \\ 0 & v_3^{(h)}  \\ 0  & -v_4^{(h)}  \\
		-v_1^{(h)} & 0  \end{array} \right ] \end{split}
\]
and $v^{(h)}$, $h=1,\ldots,16$, are the vertices of the hypercube $[-1,1]^4$.
It turns out that the minimization problem (\ref{SG_relaxedLMI2}) is feasible up to $b=b^*\approx0.841$, which is obtained for
\begin{equation}
\label{P12diag}
		P_{12} = \left[\begin{array}{cccc}
			0.595 &    0   &    0   &   0  \\
			0     &  0.595 &    0   &   0  \\
			0     &    0   &  0.595 &   0  \\
			0     &    0   &    0   & 0.595    
		\end{array}\right]  ~.
\end{equation}
In the case of Theorem \ref{nonlinearsgt2}, since $\gamma_1=\gamma_2=1/\sqrt{2}b$, the maximum gain $\gamma_{12}(b)$ allowed by the small-gain condition (\ref{sgt3}) as a function of the parameter $b$ is given by $\gamma_{12}(b)=b$. Therefore, we want to solve the following minimization problem 
\begin{align} 
\label{SG_lmi3}
	& \min_{b \geq 0, P_{12} = P_{12}^T }  b \\
	& \textrm{subject to} \notag\\
	& \left[\begin{array}{cc}
		A^T P_{12} +P_{12} A + I & P_{12} G^{(h)} \\
		G^{(h)T} P_{12} & - \gamma_{12}^2(b) I
	  \end{array}\right] \leq 0, \, h=1,\ldots,16 \notag\\ 
	& P_{12} \geq  0 \notag
\end{align}
It turns out that also problem (\ref{SG_lmi3}) is feasible up to $b=b^*\approx0.841$, and it is achieved for  the same $P_{12}$ in (\ref{P12diag}). It is worth notice that, differently from the previous example, in the case of the Thomas' system of the fourth dimension conditions (\ref{relaxedsmallgNL}) and (\ref{sgt3}) provide the same results.

As in the previous case, we compare the value of $b^*$ obtained with the small-gain conditions with the one provided by means of direct optimization. This latter minimization problem assumes the following form:
\begin{align} 
\label{SG_lmi4}
	& \min_{b \geq 0, P = P^T }  b \\
	& \textrm{subject to} \notag\\
	& J_h^{[2]}(b)^T P +P J_h^{[2]}(b) \le 0 , \, h=1,\ldots,16 \notag\\ 
	& P \geq I \notag
\end{align}
where
\[
J_h^{[2]}(b) = \left[\begin{array}{cccccc} -2\,b & v_3^{(h)} & 0 & 0 & 0 & 0\\ 0 & -2\,b & v_4^{(h)} & v_2^{(h)} & 0 & 0\\ 0 & 0 & -2\,b & 0 & v_2^{(h)} & 0\\ 0 & 0 & 0 & -2\,b & v_4^{(h)} & 0\\ -v_1^{(h)} & 0 & 0 & 0 & -2\,b & v_3^{(h)}\\ 0 & -v_1^{(h)} & 0 & 0 & 0 & -2\,b \end{array}\right] ~.  
\]
It turns out that problem (\ref{SG_lmi4}) is feasible for all $b>0.5$, and that the minimum value of $b$ is achieved for the following matrix:
\[
P = \left[\begin{array}{cccccc}
    7.27 &        0  &       0  &       0   &      0 &        0\\
         0 & 140.26  &       0   &      0   &      0  &       0\\
         0 &        0 & 128.39   &      0   &      0  &       0\\
         0 &        0  &       0   & 7.31   &      0 &        0\\
         0 &        0 &        0   &      0   & 6.53 &        0\\
         0 &        0 &        0  &       0   &      0 & 130.33\\
\end{array}\right] ~.
\]

\subsection{Thomas' example of dimension 3}
\begin{figure}[tb]
    \centering
    \includegraphics[width=0.66\columnwidth]{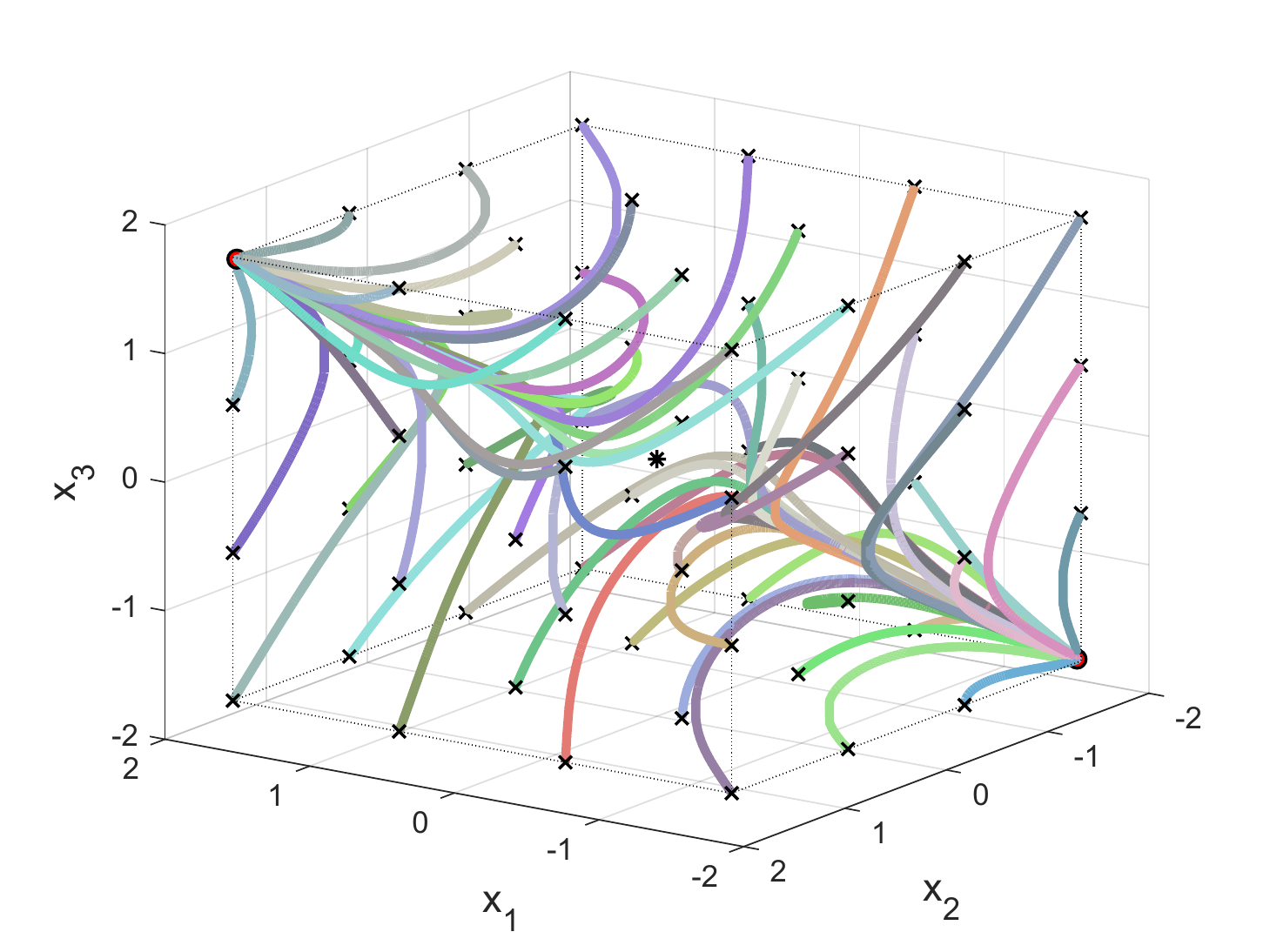}
    \caption{Evolution of trajectories started within the forward invariant set represented by the cube $[-1/b,1/b]^3$ for $b=0.58$. The crosses denote the initial conditions, while the asterisk is the unstable equilibrium point. All the other solutions converges to one of the the two stable fixed points contained in the cube.
    \label{thomas3}}
\end{figure}
In order to clarify Remark~\ref{Rem_n3}, we consider the Thomas' system of the third order. Its equations read
\[  \begin{array}{ccc}
 \dot{x}_1 &=& - b x_1 + \sin(x_2) \\
 \dot{x}_2 &=& - b x_2 + \sin(x_3) \\
 \dot{x}_3 &=& - b x_3 + \sin(x_1) 
\end{array}~,
\] 
the cube $[-1/b,1/b]^3$ is a forward invariant set, and linearization yields a second additive compound of the Jacobian of the following form:
\begin{equation}
J^{[2]}(x) =\left[
\begin{array}{ccc} 
-2\,b & \cos(x_3) & 0 \\ 
0 & -2\,b & \cos(x_2) \\ 
-\cos(x_1) & 0 & -2\,b 
\end{array}\right] ~.
\end{equation}
Since $\cos(x_i) \in [-1,1]$, then $J^{[2]}(x)\in\textrm{conv}(\mathcal{V})$, where $\textrm{conv}(\mathcal{V})$ is the interval matrix
\begin{equation}
\textrm{conv}(\mathcal{V}) =
\left[
\begin{array}{ccc} 
-2\,b & [-1,1] & 0 \\ 
0 & -2\,b & [-1,1] \\ 
{[}-1,1] & 0 & -2\,b 
\end{array}\right] ~.
\end{equation}
The modular version looks like:
\begin{equation}
\begin{array}{rcl}
\dot{\delta}_1 & = & -2 b \delta_1 + [\cos(x_3),0] \, \delta_{12} \\ 
\dot{\delta}_{12} &=& \left [ \begin{array}{cccc} -2b & \cos(x_2) \\ 0 & -2b  \end{array} \right ] \delta_{12} + \left [\begin{array}{c} 0 \\  -\cos(x_1) \end{array} \right ] \, \delta_1 \\
\end{array} ~.
\end{equation}
The gain $\gamma_1$ can be easily computed, obtaining $\gamma_1=1/(2b)$.
Then, the minimum value of $b$ allowed by the small-gain condition~(\ref{relaxedsmallgNL}) can be found by solving the minimization problem
\begin{align} 
\label{SG_lmiRem}
	& \min_{ \tilde{\eta}_1, P = P^T }  \gamma_1^2\tilde{\eta}_1 < 1 \\
	& \textrm{subject to} \notag\\
	& \left[\begin{array}{cc} 
		A_h^T P_{12} +P A_h + I & P_{12} G_h \\
		G_h^T P_{12} & - \tilde{\eta}_1 I
	  \end{array}\right] \leq 0, ~ h=1,\ldots,4 \notag\\ 
	& P_{12} = P_{12}^T \geq \varepsilon I \notag\\
	& \tilde{\eta}_1 \geq 0 \notag
\end{align}
where $\varepsilon=0.01$, 
\[
A_h = \left [ \begin{array}{cccc} -2b & v_2^{(h)} \\ 0 & -2b  \end{array} \right ], \qquad   G_h = \left [ \begin{array}{cc} 0 \\ -v_1^{(h)} \end{array} \right ] 
\]
and $v^{(h)}$, $h=1,\ldots,4$, are the vertices of the square $[-1,1]^2$, which is the projection of the forward invariant set in the space of the chosen subsystem.
It turns out that problem (\ref{SG_lmiRem}) is feasible for all $b>0.575$, and that the optimal matrix $P$ providing the minimum value of $b$ is
\[
P_{12} =\left[
\begin{array}{cc} 
    0.87		&	0 \\
    0		&	1.52 
\end{array}\right] ~.  
\]
As in the case of the fourth order Thomas' system, it can be proved that conditions (\ref{relaxedsmallgNL}) and (\ref{sgt3}) provide the same results.
In Fig.~\ref{thomas3} the convergence to two distinct fixed points of the trajectories started within the forward invariant set is illustrated by numerical simulations obtained for $b=0.58$.

It is interesting also in this case to compare the value of $b^*$ provided by the small-gain condition with the one achievable by considering the second additive compound $J^{[2]}(x)$. The minimization problem becomes
\begin{align} 
\label{SG_lmi5}
	& \min_{ b \geq 0, P = P^T }  b \\
	& \textrm{subject to} \notag\\
	& J_h^{[2]}(b)^T P +P J_h^{[2]}(b)  \leq 0  , h=1,\ldots,4 \notag\\ 
	& P \geq I \notag
\end{align}
where
\[
J_h^{[2]}(b) =\left[
	\begin{array}{ccc} 
		-2\,b & v_3^{(h)} & 0 \\ 
		0 & -2\,b & v_2^{(h)} \\ 
		-v_1^{(h)} & 0 & -2\,b 
	\end{array}\right] ~.  
\]
and $v^{(h)}$, $h=1,\ldots,8$, are the vertices of the cube $[-1,1]^3$.
We get that problem (\ref{SG_lmi5}) is feasible for all $b>0.44$ and the optimal matrix $P$ corresponding the minimum $b$ is the identity matrix.

\section{Conclusions}
This paper proposes sufficient small-gain conditions for assessing non-oscillatory behavior of solutions of feedback interconnected systems.
The general criterion is based on the notion of $2$-contraction and provides a modular approach for the stability analysis of second additive compound matrix variational equations, arising by considering virtual displacements for linear or nonlinear dynamical systems.
The conditions are expressed as functionals that must be less than unity and which are computed once the system has been divided into interconnected subsystems.
The proposed functionals account both for the individual gains of each subsystem's second additive variational equations, and for an ``interconnection'' gain, which arises from considerations on the Kronecker's sum of the Jacobians of the individual subsystems.
This approach opens the way for further extensions in several directions, such as modular approaches for $2$-contraction of large-scale interconnection systems or modular approaches for general $k$-contraction. 

\section*{Appendix}
\subsection*{Proof of Theorem 1}

As a preliminary observation, notice that the LMI in~(\ref{relaxedsmallg}) is just a more convenient formulation of~(\ref{eq:gamma12 LMI}), where $\eta_1^2$  and $\eta_2^2$ have been replaced with $\tilde{\eta}_1$  and $\tilde{\eta}_2$ in order to keep the matrix inequality linear with respect to its unknowns.
The proof is based on the construction of a block-diagonal quadratic Lyapunov function, exploiting the equivalent formulation of  $A^{[2]}$ dynamics provided by equation~(\ref{coupledsystem}).
To this end, notice that, after a suitable reordering of state-variables, the matrix $A^{[2]}$ can be transformed as:
\[  \mathcal{A} = \left [  \begin{array}{ccc}  A_{11}^{[2]} & B_1 & 0 \\ G_1 & A_{11} \oplus A_{22}  & G_2 \\
	0 & B_2 & A_{22}^{[2]} 
\end{array}  \right ].       \]
We consider a quadratic Lyapunov function defined by the following symmetric definite matrix:
\[ \mathcal{P} =  \left [  \begin{array}{ccc}  \lambda_1 P_1 & 0 & 0 \\ 0 & P_{12}  & 0 \\
	0 & 0 & \lambda_2 P_2 
\end{array}  \right ] . \]
Matrices $P_1$ and $P_2$ are as in~(\ref{gamma1}) and~(\ref{gamma2}) and $\lambda_1>0$, $\lambda_2>0$ are to be chosen later.
Direct calculations show that $\mathcal{A}^T \mathcal{P} + \mathcal{P} \mathcal{A}$ gets the formulation reported in~(\ref{Th2 - ATP+PA}) and it satisfies the inequalities~(\ref{Lyap_Th2 - 1}) and~(\ref{Lyap_Th2 - 2}) by virtue of LMIs~(\ref{gamma1}) and~(\ref{gamma2}).
\begin{align}
	\label{Th2 - ATP+PA}
	\mathcal{A}^T \mathcal{P} + \mathcal{P} \mathcal{A} &=
	\left [  \begin{array}{ccc}   \lambda_1 \big( { A_{11}^{[2]}}^T P_1 + P_1 A_{11}^{[2]} \big)  & \lambda_1 P_1 B_1 + G_1^T P_{12} & 0 \\
		\lambda_1 B_1^T P_1 + P_{12} G_{1} &   (A_{11} \oplus A_{22})^T P_{12} + P_{12} ( A_{11} \oplus A_{22} ) & \lambda_2 B_2^T P_2 + P_{12}  G_2 \\
		0 & \lambda_2 P_2 B_2 + G_2^T P_{12} &  \lambda_2 \big( {A_{22}^{[2]}}^T P_2 + P_2 A_{22}^{[2]} \big) 
	\end{array} \right ]\\
	\label{Lyap_Th2 - 1}
	&\leq \left [  \begin{array}{ccc}   -\lambda_1 I &  G_1^T P_{12} & 0 \\
		P_{12} G_{1} & \lambda_1 \gamma_1^2 I +  (A_{11} \oplus A_{22})^T P_{12} + P_{12} ( A_{11} \oplus A_{22} ) & \lambda_2 B_2^T P_2 + P_{12}  G_2 \\
		0 & \lambda_2 P_2 B_2 + G_2^T P_{12} &  \lambda_2 \big( {A_{22}^{[2]}}^T P_2 + P_2 A_{22}^{[2]} \big) 
	\end{array} \right ]\\
	\label{Lyap_Th2 - 2}
	&\leq \left [  \begin{array}{ccc}   - \lambda_1 I &   G_1^T P_{12} & 0 \\
		P_{12} G_{1} & (\lambda_1 \gamma_1^2 + \lambda_2 \gamma_2^2) I +  (A_{11} \oplus A_{22})^T P_{12} + P_{12} ( A_{11} \oplus A_{22} ) &   P_{12}  G_2 \\
		0 &   G_2^T P_{12} & -\lambda_2 I
	\end{array} \right ]
\end{align}
The inequality~(\ref{Lyap_Th2 - 2}) can be rearranged via a suitable permutation matrix $\mathcal{S}$, so that the same inequality assumes the form:
\begin{equation}
	\mathcal{S}^T\left(\mathcal{A}^T \mathcal{P} + \mathcal{P} \mathcal{A}\right)\mathcal{S} \leq
 	\left [  \begin{array}{ccc}  
 		(\lambda_1 \gamma_1^2+ \lambda_2 \gamma_2^2) I + (A_{11} \oplus A_{22})^T P_{12} + P_{12} ( A_{11} \oplus A_{22} ) &  P_{12} G_1 & P_{12} G_2 \\
 		G_1^T P_{12} & -\lambda_1 I & 0 \\
 		G_2^T P_{12} & 0 & - \lambda_2 I
 	\end{array} \right ]
 	\label{Th2_Rearr}
\end{equation} 
Finally, if we select $P_{12}=P_{12}^*$ and $\lambda_1=\tilde{\eta}_1^* + \sigma$, $\lambda_2=\tilde{\eta}_2^* +\sigma$ for some $\sigma >0$ to be chosen later, where $P_{12}^*$, $\tilde{\eta}_1^*$,  $\tilde{\eta}_2^*$ solve the small-gain condition (\ref{relaxedsmallg}), we get 
\begin{equation}
 	\mathcal{S}(\mathcal{A}^T \mathcal{P} + \mathcal{P} \mathcal{A})\mathcal{S} \leq
 	\left [  \begin{array}{ccc}  
 		-\alpha I  &  0 & 0 \\
 		0 & - \sigma I & 0 \\
 		0 & 0 & - \sigma I
 	\end{array} \right ]
 	\label{Lambda_Th2}
\end{equation}
with
\begin{align*}
	\alpha = 1-\tilde{\eta}_1^* \gamma_1^2+ \tilde{\eta}_2^* \gamma_2^2 - \sigma ( \gamma_1^2 + \gamma_2^2) .
\end{align*}
If we choose $\sigma$ such that
\begin{align*}
\sigma < \frac{ 1 -  (\tilde{\eta}^* \gamma_1^2+ \tilde{\eta}^* \gamma_2^2) }{ \gamma_1^2 + \gamma_2^2 } ,
\end{align*}
we get $\alpha>0$, which by virtue of (\ref{Lambda_Th2}) implies that $\mathcal{A}$ is Hurwitz, thus proving 2-contraction of the matrix $A^{[2]}$.

\subsection*{Proof of Theorem 2}
The argument proceeds along the same lines as in the proof of Theorem~\ref{firstmain} by constructing a quadratic Lyapunov function defined by the following symmetric definite matrix:
\[ \mathcal{P} =  \left [  \begin{array}{ccc}  P_1 & 0 & 0 \\ 0 & \lambda P_{12}  & 0 \\
	0 & 0 & P_2 
\end{array}  \right ], \]
where $P_1$, $P_2$, $P_{12}$ are as in (\ref{gamma1}), (\ref{gamma2}), (\ref{gamma12}) and $\lambda>0$ is to be chosen later.
A direct calculation leads to the formulation of $\mathcal{A}^T \mathcal{P} + \mathcal{P} \mathcal{A}$ as reported in~(\ref{thrm2 - ATP+PA}), which satisfies condition (\ref{thrm2 - 1}) and (\ref{thrm2 - 2}) thanks to the LMIs (\ref{gamma1}) and (\ref{gamma2}).
\begin{align}   
	\label{thrm2 - ATP+PA}
	\mathcal{A}^T \mathcal{P} + \mathcal{P} \mathcal{A} =&
	\left [  \begin{array}{ccc}   {A_{11}^{[2]}}^T P_1 + P_1 A_{11}^{[2]} & P_1 B_1 + \lambda G_1^T P_{12} & 0 \\
		B_1^T P_1 + \lambda P_{12} G_{1} &  \lambda \big( (A_{11} \oplus A_{22})^T P_{12} + P_{12} ( A_{11} \oplus A_{22} )\big)  & B_2^T P_2 + \lambda P_{12}  G_2 \\
		0 & P_2 B_2 + \lambda G_2^T P_{12} & {A_{22}^{[2]}}^T P_2 + P_2 A_{22}^{[2]}
	\end{array} \right ]\\
	\label{thrm2 - 1}
	\leq& \left [  \begin{array}{ccc}   -I &  \lambda G_1^T P_{12} & 0 \\
		\lambda P_{12} G_{1} & \gamma_1^2 I + \lambda \big( (A_{11} \oplus A_{22})^T P_{12} + P_{12} ( A_{11} \oplus A_{22} )\big)  & B_2^T P_2 + \lambda P_{12}  G_2 \\
		0 & P_2 B_2 + \lambda G_2^T P_{12} & {A_{22}^{[2]}}^T P_2 + P_2 A_{22}^{[2]}
	\end{array} \right ] \\
	\label{thrm2 - 2}
	\leq& \left [  \begin{array}{ccc}   -I &  \lambda G_1^T P_{12} & 0 \\
		\lambda P_{12} G_{1} & (\gamma_1^2+ \gamma_2^2) I + \lambda \big( (A_{11} \oplus A_{22})^T P_{12} + P_{12} ( A_{11} \oplus A_{22} )\big)  &  \lambda P_{12}  G_2 \\
		0 &  \lambda G_2^T P_{12} & -I
	\end{array} \right ]
\end{align}
The last matrix in~(\ref{thrm2 - 2}) can be properly rearranged through suitable permutations of the state variables via a permutation matrix $\mathcal{S}$ to get the formulation reported in~(\ref{Matrixgamma2}), which finally leads to inequality~(\ref{LMIFin}) by exploiting~LMI (\ref{gamma12}).
\begin{align}
	\mathcal{S}^T\left(\mathcal{A}^T \mathcal{P} + \mathcal{P} \mathcal{A}\right)\mathcal{S} \leq&
	\left[\begin{array}{cc}  
	(\gamma_1^2+ \gamma_2^2) I + \lambda \big( (A_{11} \oplus A_{22})^T P_{12} + P_{12} ( A_{11} \oplus A_{22} ) \big)  &  \lambda P_{12} [G_1, G_2] \\
	\lambda [ G_1,G_2]^T P_{12} & -I
	\end{array}\right]
	\label{Matrixgamma2} \\
	\leq&  \left [  \begin{array}{cc}  
		(\gamma_1^2+ \gamma_2^2 - \lambda) I   &  0 \\
		0 & (\lambda \gamma_{12}^2 -1)I
	\end{array} \right ]
	\label{LMIFin}
\end{align}
Hence, from
\begin{align*}
	\mathcal{S}^T (  \mathcal{A}^T \mathcal{P} + \mathcal{P} \mathcal{A})\mathcal{S} \leq
	\left[\begin{array}{cc}  
	(\gamma_1^2+ \gamma_2^2) I - \lambda I  &  0 \\
	0 & \lambda \gamma_{12}^2 I - I
	\end{array}\right]
\end{align*}
one gets that $\mathcal{A}^T \mathcal{P} + \mathcal{P} \mathcal{A}<0$, if $\lambda$ is chosen so that:
\begin{align*}
	\gamma_1^2 + \gamma_2^2 < \lambda < \frac{1}{\gamma_{12}^2}.
\end{align*}

\subsection*{Proof of Theorem 3}
To see the result, notice that the variational equation~(\ref{standardvar}) can be rearranged through suitable permutations according to~(\ref{modularvariational}).
In particular,
\begin{align*}
\dot{ \delta} = \mathcal{A} (x) \delta, 
\end{align*}
for the block matrix
\begin{align*}
	\mathcal{A}(x) = \left [ \begin{array}{ccc} J_{11}^{[2]} (x) & B_1(x) & 0 \\
	G_1(x) & J_{11}(x) \oplus J_{22}(x) & G_2 (x) \\
	0 & B_2(x) & J_{22}^{[2]} (x) 
\end{array} \right ].
\end{align*}

We adopt a candidate solution for~(\ref{contractionLMI}) of the following form:
\begin{align*}
	\mathcal{P}(x) =   \left[\begin{array}{ccc}
	\lambda_1 P_1(x) & 0 & 0 \\
	0 & P_{12} & 0 \\
	0  & 0 & \lambda_2 P_2(x)
	\end{array} \right ].
\end{align*}
A direct computation shows that $\mathcal{A}^T(x) \mathcal{P} + \mathcal{P} \mathcal{A}(x) +\dot{\mathcal{P}}(x)$ assumes the form reported in~(\ref{LyapEqPdot1}).
\begin{align}
	\notag
	&\mathcal{A}^T(x)\mathcal{P}(x) + \mathcal{P}(x)\mathcal{A}(x) + \dot{\mathcal{P}}(x) \\
	&= \left [\begin{array}{ccc}
		\lambda_1 \big({J_{11}^{[2]}}^T(x) P_1(x) +P_1(x) J_{11}^{[2]}(x)\big)
			& \lambda_1 P_1(x) B_1(x) + G_1^T(x) P_{12}
			& 0 \\
		+\lambda_1 \dot{P}_1(x) && \\
		\lambda_1 B_1^T(x) P_1(x) + P_{12} G_{1}(x)
			&  \big(J_{11}(x) \oplus J_{22}(x)\big)^T P_{12}
			& \lambda_2 B_2^T(x) P_2(x) + P_{12}  G_2(x) \\
		&  + P_{12} \big(J_{11}(x) \oplus J_{22}(x)\big) & \\
		0 
			& \lambda_2 P_2(x) B_2(x) + G_2^T(x) P_{12}
			& \lambda_2\big({A_{22}^{[2]}}^T(x) P_2(x) + P_2(x) A_{22}^{[2]}(x)\big) \\
		&& +\lambda_2 \dot{P}_2(x)
	\end{array} \right ] 
	\label{LyapEqPdot1}
\end{align}
%
%
Then, the proof follows along similar lines as the proof of Theorem~\ref{firstmain} by applying the inequalities considered in (\ref{j1gain}), (\ref{j2gain}) and (\ref{hull_relaxed_cond}), this latter evaluated in the vertexes of the hull.

\subsection*{Proof of Theorem 4}
The argument proceeds along the same lines as in the proof of Theorem~\ref{secondmain}, by constructing a quadratic Lyapunov function defined by the following symmetric definite matrix:
\begin{align}
	\mathcal{P}(x) =  \left[\begin{array}{ccc}
	P_1(x) & 0 & 0 \\
	0 & \lambda P_{12} (x) & 0
	\\ 0  & 0 & P_2 (x)
	\end{array}\right] ~.
\end{align}
Direct computations lead $\mathcal{A}^T(x) \mathcal{P}(x) + \mathcal{P}(x) \mathcal{A}(x) + \dot{\mathcal{P}}(x)$ to the formulation reported in~(\ref{LyapEqPdot}).
\begin{align}
	& \mathcal{A}^T(x) \mathcal{P}(x) + \mathcal{P}(x) \mathcal{A}(x) + \dot{\mathcal{P}}(x) \notag \\
	&= \left [\begin{array}{ccc}
	\rule{0pt}{1.5em}{\phantom{A}}
	{J_{11}^{[2]}}^T(x) P_1(x) + P_1(x) J_{11}^{[2]}(x)
		& P_1(x) B_1(x) + \lambda G_1^T(x) P_{12}(x)
		& 0 \\
	\rule[-0.5em]{0pt}{0.5em}{\phantom{A}}
	+\dot{P}_1(x) & & \\
	B_1^T(x) P_1(x) + \lambda P_{12}(x) G_{1}(x)
		& \rule{0pt}{1.5em}{\phantom{A}} \lambda(J_{11}(x) \oplus J_{22}(x))^T P_{12}(x)
		& B_2^T(x) P_2(x) + \lambda P_{12}(x)  G_2(x) \\
	& \rule[-0.5em]{0pt}{0.5em}{\phantom{A}} +\lambda P_{12}(x) ( J_{11}(x) \oplus J_{22}(x) )+\lambda\dot{P}_{12}(x) & \\
	0
		& P_2(x) B_2(x) + \lambda G_2^T(x) P_{12}(x)
		& \rule{0pt}{1.5em}{\phantom{A}}
		 {A_{22}^{[2]}}^T(x) P_2(x) + P_2(x) A_{22}^{[2]}(x) \\
	& &  \rule[-0.5em]{0pt}{0.5em}{\phantom{A}} +\dot{P}_2(x)
	\end{array} \right ] 
	\label{LyapEqPdot}
\end{align}
The proof follows along similar lines as the proof of Theorem~\ref{secondmain} by applying the inequalities considered in (\ref{j1gain}), (\ref{j2gain}) and (\ref{j12gain}).

\bibliographystyle{unsrt}
\bibliography{main_1_single_column}

\end{document}